\documentclass[a4paper,twoside,english]{amsart}
\usepackage[T1]{fontenc}
\usepackage[latin1]{inputenc}
\usepackage{amssymb}

\makeatletter

\newcommand{\noun}[1]{\textsc{#1}}

 \theoremstyle{plain}    
 
 \numberwithin{equation}{section}
 \numberwithin{figure}{section}
 \theoremstyle{plain}
 \theoremstyle{remark}    
 \newtheorem*{acknowledgement*}{Acknowledgment} 

\usepackage{babel}
\makeatother
\begin{document}
ULM-TP{/}05-4

October 2005

\vspace{2 cm}
\title{On the characteristic exponents of Floquet solutions to the Mathieu
equation}

\author{Jan Eric Str{\"a}ng}

\address{Abteilung Theoretische Physik, Universit{\"a}t Ulm, Albert-Einstein
Allee 11, D89069 Ulm, Germany}
\email{eric.straeng@uni-ulm.de}
\begin{abstract}
We study the Floquet solutions of the Mathieu equation. In order
to find an explicit relation between the characteristic exponents
and their corresponding eigenvalues of the Mathieu operator, we consider 
the Whittaker-Hill formula. This gives an explicit relation between 
 the eigenvalue and its characteristic exponent. The equation is explicit 
up to a determinant of an infinite dimensional matrix. We find a third-order 
linear recursion for which this determinant is exactly the limit. 
An explicit solution for third-order linear recursions is obtained 
which enables us to write the determinant
explicitly.
\end{abstract}
\maketitle

\section{Introduction\label{par:Introduction}}

The Mathieu equation is a special case of Hill's equation \[
\frac{d^{2}\psi}{dz^{2}}+\left(a-2q\phi\left(z\right)\right)\psi=0\]
where $\phi\left(z\right)$ is a $\pi$-periodic differentiable function
of the real variable $z$ of maximum absolute value 1 and $a,\, q$ are
real constants%
\footnote{Throughout this paper we will make an effort to use existing conventions
regarding the notations of the Hill and Mathieu equations (see for
example \cite{McLachlan})%
}. Thus, the equation can be written in terms of the Fourier transform
of $\phi\left(z\right)$ as\begin{equation}
\frac{d^{2}\psi}{dz^{2}}+\left(\theta_{0}+2\sum_{r=1}^{\infty}\theta_{2r}\cos\left(2rz\right)\right)\psi=0\label{eq:Hill}\end{equation}
if the function $\phi\left(z\right)$ is even. The Mathieu equation
is a restriction to $\phi\left(z\right)=\cos\left(2z\right)$ of the
Hill equation, i.e.\begin{eqnarray}
\frac{d^{2}\psi}{dz^{2}}+\left(a-2q\cos\left(2z\right)\right)\psi & = & 0.\label{eq:Mathequ}\end{eqnarray}

The equation was originally studied in the context of vibrational
modes of an elliptic membrane by \noun{{\'E}.} Mathieu \cite{Mathieu}.
Considering time periodic solutions, the initial wave equation can
be separated and written in elliptic coordinates \cite{McLachlan,Meixner},
consequently the wave equation takes the form of eq. (\ref{eq:Mathequ}).
In a similar fashion, this procedure can be applied to heat equations
on elliptic domains \cite{McLachlan}. Furthermore, the equation appears
in solid mechanics with a periodic perturbation, e.g. for a rod fixed
on one end and being under periodic tension at the other end \cite{McLachlan}.

The Hill equation also appears in quantum mechanics as an eigenvalue
problem for a particle in a periodic potential \cite{Kohn}. If we
restrict the potential to $2Q\cos\left(2z\right)$ in one dimension,
the Schr{\"o}dinger equation $\left(\hbar\equiv1,\,2m\equiv1\right)$\[
\imath\frac{\partial\psi}{\partial t}=-\frac{\partial^{2}\psi}{\partial z^{2}}+2Q\cos\left(2z\right)\psi\]
will lead to an eigenvalue problem for the operator $\hat{H}=-\partial^{2}/\partial z^{2}+2Q\cos\left(2z\right)$,\[
E\psi=-\frac{\partial^{2}\psi}{\partial z^{2}}+2Q\cos\left(2z\right)\psi,\]
which is easily written in the form of eq. (\ref{eq:Mathequ})\begin{equation}
\frac{\partial^{2}\psi}{\partial z^{2}}+\left(E-2Q\cos\left(2z\right)\right)\psi=0,\label{eq:SSP}\end{equation}
identifying $a=E$ and $q=Q$. This particular problem is of special
importance in solid state physics where the crystal lattice atoms
constitute a periodic potential for loosely bound electrons. Eq. (\ref{eq:SSP})
is a crude one dimensional simplification of such a potential.

After a short introduction to the Mathieu equation, we will describe
(section \ref{sec:Floquet-Solutions-and}) the Floquet theorem and
its application to the Mathieu equation. We will then state some general
properties of the Floquet solutions (section \ref{sec:Determinental-equation})
and more specifically the Whittaker-Hill formula (section \ref{sec:Whittaker-Hill-formula})
which has given rise to this work. We will then focus on the calculation
of an infinite dimensional determinant (section \ref{sec:Calculating})
needed to solve the equation given by the Whittaker-Hill formula.
In section \ref{recursol} we describe a method to obtain an explicit
result for a third order linear recursion. In section \ref{sec:An-explicit-form},
an explicit form of this determinant is given.

\section{The Mathieu equation}

There exist a multitude of solutions to the Mathieu equation (\ref{eq:Mathequ}),
both periodic and non-periodic. What is distinctive is that they cannot
generally be written in terms of elementary functions, i.e., they
are transcendental functions. For a summary see, e.g., \cite{Gradsteyn,Abrahamovitz}
and for a more complete account with examples see \cite{McLachlan,Meixner}. There 
also exists a number of approximation schemes for the Floquet solutions 
\cite{Gradsteyn,Abrahamovitz,McLachlan,Meixner}. To our knowledge, the most 
recent approximations are due to Frenkel and Portugal \cite{Frenkel}.
We will restrict our study to periodic solutions and more specifically
to the Floquet solutions of eq. (\ref{eq:Mathequ}).

\subsection{\label{sec:Floquet-Solutions-and}Floquet Solutions}

As the coefficients of the Mathieu eq. (\ref{eq:Mathequ}) are periodic,
one can apply the Floquet theorem. According to Floquet \cite{Floquet},
a differential equation with periodic coefficients admits at least
one periodic solution of the second kind, i.e., if $\psi\left(z\right)$
is a uniformly continuous solution, then \begin{equation}
\psi\left(z+\tau;\, a,\, q\right)=\varepsilon\psi\left(z;\, a,\, q\right),\,\forall z\in\mathbb{R},\label{eq:secondkind}\end{equation}
where $\tau$ is the period of the coefficients and the multiplier
$\varepsilon$ is a constant. If $\varepsilon=1$ the solution is
of the same period as the coefficients and is called periodic of the
first kind. Further identification of the multiplier leads to the
solutions of the form\begin{equation}
\psi\left(z;\, a,\, q\right)=e^{\imath\nu z}u_{\nu}\left(z;\, a,\, q\right),\,\nu\in\mathbb{R},\label{eq:Floquetsolution}\end{equation}
where $u_{\nu}\left(z\right)$ is a periodic and uniformly continuous
function of the first kind with the same period as the coefficient.
We will call such solutions \emph{Floquet solutions} to the Mathieu equation,
and call $\nu$ its \emph{characteristic exponent}.

It should be pointed out that though the sum of Floquet solutions
is a solution to eq. (\ref{eq:Mathequ}), it is no longer a Floquet
solution. Furthermore, the solutions according to eq. (\ref{eq:Floquetsolution})
with real $\nu$ constitute so-called stable solutions. There exists
values $\left(a,\, q\right)$ for which such solutions do not exist,
in which case $\nu$ becomes complex. These solutions are no longer
bounded and hence are called unstable. The eigenvalues $a$ for which
$\nu$ is real form bands \cite{SimonReed}, with gaps corresponding
to complex characteristic exponents. For non integer characteristic
exponent, there exists a space of solutions spanned by two independent
solutions \cite{Kohn,McLachlan,Meixner,SimonReed}. For integer values
of $\nu$ there exists only one solution which is usually written
in terms of the Mathieu function $\mathrm{me}_{\nu}$ \cite{Meixner,Abrahamovitz,Gradsteyn,McLachlan}.
Integer $\nu$ are border solutions to the eigenvalue bands.

We will state an important connection between the eigenvalue and the
characteristic exponent as given by Kohn \cite{Kohn}. The solutions
$\psi_{1}$, $\psi_{2}$ to a given pair $\left(a,\, q\right)$ with
initial conditions \begin{eqnarray*}
\psi_{1}\left(0;\, a,\, q\right)=1, &  & \psi_{1}^{\prime}\left(0;\, a,\, q\right)=0\\
\psi_{2}\left(0;\, a,\, q\right)=0, &  & \psi_{2}^{\prime}\left(0;\, a,\, q\right)=1\end{eqnarray*}
form a fundamental system, i.e., they are linearly independent \cite{Meixner}.
Any solution to a pair $\left(a,\, q\right)$ can be written with
those solutions. A Floquet solution satisfies\[
\psi\left(z+\tau;\, a,\, q\right)=\varepsilon\psi\left(z;\, a,\, q\right),\]
hence\begin{eqnarray*}
\psi\left(\tau;\, a,\, q\right) & = & \varepsilon\psi\left(0;\, a,\, q\right),\\
\psi^{\prime}\left(\tau;\, a,\, q\right) & = & \varepsilon\psi^{\prime}\left(0;\, a,\, q\right),\end{eqnarray*}
where $\varepsilon=e^{\imath\nu\tau}$. Rewriting $\psi\left(z;\, a,\, q\right)$
as a linear combination of $\psi_{1}\left(z;\, a,\, q\right)$ and
$\psi_{2}\left(z;\, a,\, q\right)$\[
\psi=A\psi_{1}+B\psi_{2}\]
one obtains the equation\begin{equation}
\varepsilon^{2}-2\mu\left(a\right)\varepsilon+1=0\label{eqdiscr}\end{equation}
upon elimination of $A$ and $B$. Here $\mu\left(a\right)$ is an
entire function of $a$ that can be expressed as \[
\mu\left(a\right)=\frac{1}{2}\left(\psi_{1}\left(\tau;\, a,\, q\right)+\psi_{2}^{\prime}\left(\tau;\, a,\, q\right)\right).\]
Insertion of $\varepsilon=e^{\imath\nu\tau}$ according to the Floquet
theorem in eq. (\ref{eqdiscr}) and subsequent multiplication by $e^{-\imath\nu\tau}$
gives\begin{equation}
\cos\left(\nu\tau\right)=\mu\left(a\right).\label{eq:Kohndisp}\end{equation}
This equation can be used to obtain a relation between the characteristic
exponents and the eigenvalues of the operator $\hat{H}$. Unfortunately
the function $\mu\left(a\right)$ cannot be written in such a way
as to solve eq. (\ref{eq:Kohndisp}).

\subsection{\label{sec:Determinental-equation}Determinantal equation}

Since $u_{\nu}$ is periodic, we will seek solutions $u_{\nu}\in L_{2}\left(\left[0,\,\pi\right]\right)$
and extend these solution to $z\in\mathbb{R}.$ We may expand the
entire solution as the Fourier series \[
\psi\left(z,\, a\right)=e^{\imath\nu z}\sum_{\kappa\in\mathbb{Z}}c_{2\kappa}\left(\nu;\, a,\, q\right)e^{2i\kappa z},\]
which, inserted into eq. (\ref{eq:Mathequ}), gives the recurrence
relation\begin{equation}
\left(\left(2\kappa-\nu\right)^{2}-a\right)c_{2\kappa}+q\left(c_{2\left(\kappa+1\right)}+c_{2\left(\kappa-1\right)}\right)=0,\,\forall\kappa\in\mathbb{Z}\label{eq:recursionfourier}\end{equation}
 for the Fourier coefficients $c_{2\kappa}\left(a,\,\nu\right)$.
Since $u_{\nu}\in L_{2}\left(\left[0,\,\pi\right]\right)$, we see
that $\left\{ c_{2\kappa}\right\} _{\kappa\in\mathbb{Z}}\in\ell_{2}\left(\mathbb{Z}\right)$
whereupon\[
\sum_{\kappa\in\mathbb{Z}}\left|c_{2\kappa}\left(\nu;\, a,\, q\right)\right|^{2}<\infty.\]

For any finite (truncated) upper limit $n\in\mathbb{Z}$, eq. (\ref{eq:recursionfourier})
can be written as a matrix equation, \begin{equation}
A_{n}\left(\nu;\, a,\, q\right)\overline{c}_{n}=0.\label{eq:matrixequ}\end{equation}
where $\overline{c}_{n}=\left(c_{-2n},\ldots,\, c_{-2},\, c_{0},\, c_{2},\, c_{4},\,\ldots,\, c_{2n}\right)^{T}$
and $A_{n}\left(\nu;\, a,\, q\right)$ is a $2n+1\times2n+1$ matrix.
For the sake of convergence (see below) we will define the equivalent
system of equations\begin{eqnarray}
c_{2\kappa}+\frac{q\left(c_{2\left(\kappa+1\right)}+c_{2\left(\kappa-1\right)}\right)}{\left(2\kappa-\nu\right)^{2}-a} & = & 0,\,\forall\kappa\in\mathbb{Z},\label{eq:modifiedequfourier}\end{eqnarray}
whereupon eq. (\ref{eq:recursionfourier}) takes the form\begin{equation}
\mathcal{A}_{n}\left(\nu;\, a,\, q\right)\overline{c}_{n}=0\label{eq:eqdicrpre}\end{equation}
with\begin{equation}
\mathcal{A}_{n}\left(\nu;\, a,\, q\right)=\left(\begin{array}{ccccccccccc}
1 & \xi_{2n}\\
\xi_{2n-2} & \cdot & \cdot & \cdot & \cdot & \cdot & \cdot & \cdot & \cdot & \cdot\\
 & \cdot & \xi_{4} & 1 & \xi_{4} & 0 & 0 & 0 & 0 & \cdot\\
 & \cdot & 0 & \xi_{2} & 1 & \xi_{2} & 0 & 0 & 0 & \cdot\\
 & \cdot & 0 & 0 & \xi_{0} & 1 & \xi_{0} & 0 & 0 & \cdot\\
 & \cdot & 0 & 0 & 0 & \xi_{-2} & 1 & \xi_{-2} & 0 & \cdot\\
 & \cdot & 0 & 0 & 0 & 0 & \xi_{-4} & 1 & \xi_{-4} & \cdot\\
 & \cdot & \cdot & \cdot & \cdot & \cdot & \cdot & \cdot & \cdot & \cdot & \xi_{-2n+2}\\
 &  &  &  &  &  &  &  &  & \xi_{-2n} & 1\end{array}\right)\label{eq:matrixdarst}\end{equation}
and\[
\xi_{2\kappa}=\frac{q}{\left(2\kappa-\nu\right)^{2}-a},\,\forall\kappa\in\mathbb{Z}.\]

For the finite dimensional case, finding the non-trivial solutions
of eq. (\ref{eq:eqdicrpre}) is equivalent to demanding\[
\det\left(\mathcal{A}_{n}\right)=0.\]
The determinant can, however, not always be defined for the infinite
dimensional case.

The determinant $\det\left(\mathcal{C}+\mathbf{I}\right)$ of an operator $\mathcal{C}+\mathbf{I}$ 
defined on an infinite dimensional Hilbert space, can in certain cases
be defined by the product of $\left(1+\lambda_{i}\right)$ where $\lambda_{i}$ are
the eigenvalues of $\mathcal{C}$. For trace class operators $\mathcal{C}$, 
this product is well defined, independent of the choice of basis and
converges absolutely \cite{TraceideaSimon}. A trace class operator,
suitably defined on a separable Hilbert space, has the property\[
\mathbf{Tr}\left(\left|\mathcal{C}\right|\right)=\sum_{i=1}^{\infty}\mu_{i}<\infty,\]
where $\mu_{i}=\left|\lambda_{i}\right|$ are the singular values
of $\mathcal{C}$ \cite{TraceideaSimon,Kato}. For such operators
one can define the trace norm\[
\left\Vert \mathcal{C}\right\Vert _{\mathbf{Tr}}=\sum_{i=1}^{\infty}\mu_{i}.\]
We can consider the operator $\mathcal{B}_{n}:\,\mathbb{C}^{2n+1}\rightarrow\mathbb{C}^{2n+1}$
with $\mathcal{B}_{n}:=\mathcal{A}_{n}-\mathbf{I}$ . By letting $n\rightarrow\infty$,
the limiting matrices $\mathcal{A}$, $\mathcal{B}$ will constitute
operators on the infinite dimensional Hilbert space $\ell_{2}\left(\mathbb{Z}\right)$.
Furthermore, $\mathcal{B}$ can be written as\[
\mathcal{B}=\mathcal{B}_{+1}+\mathcal{B}_{-1}\]
where $\mathcal{B}_{+1}$ and $\mathcal{B}_{-1}$ are single sub-
and super-diagonal, respectively, with elements $\left(\ldots,\,\xi_{-2n},\,\xi_{-2n+2},\,\ldots,\,\xi_{-2},\,\xi_{0},\,\xi_{2},\,\ldots\right)$.
The singular values of these operators will hence be $\left|\xi_{2\kappa}\right|,\,\kappa\in\mathbb{Z}$.
The sum of the singular values of these operators will be\begin{eqnarray*}
\mathbf{Tr}\left(\left|\mathcal{B}_{\pm1}\right|\right) & = & \sum_{\kappa\in\mathbb{Z}}\left|\xi_{2\kappa}\right|\\
 & = & \sum_{\kappa\in\mathbb{Z}}\left|\frac{q}{\left(2\kappa-\nu\right)^{2}-a}\right|\end{eqnarray*}
which can be rewritten\begin{eqnarray}
\sum_{\kappa\in\mathbb{Z}}\left|\xi_{2\kappa}\right| & = & \frac{1}{4}\sum_{\kappa\in\mathbb{Z}}\left|\frac{1}{\kappa^{2}}\right|\cdot\left|\frac{1}{\left(1-\frac{\nu}{2\kappa}\right)^{2}-\frac{a}{4\kappa^{2}}}\right|.\label{eq:converge}\end{eqnarray}
This sum will converge since $\sum_{1}^{\infty}\frac{1}{\kappa^{\chi}},\,\chi>1$,
converges and the terms under the nominator in eq. (\ref{eq:converge})
are inferior to $\left|r\right|^{-p},\,1<p<2$ for large enough $\kappa$.
This is valid for all finite values of $a$ and $\nu$ for which no
denominator of $\xi_{2\kappa}$ vanishes, i.e., \begin{equation}
2\kappa-\nu\neq\pm\sqrt{a}.\label{eq:constranu}\end{equation}
The operators $\mathcal{B}_{+1}$ and $\mathcal{B}_{-1}$ defined
on $\ell_{2}\left(\mathbb{Z}\right)$ are therefore trace class operators
with trace norms\[
\left\Vert \mathcal{B}_{\pm1}\right\Vert _{\mathbf{Tr}}=\mathbf{Tr}\left(\left|\mathcal{B}_{\pm1}\right|\right).\]
Furthermore, by the triangle inequality of the trace norm \cite{Kato}
we can write\[
\left\Vert \mathcal{B}_{+1}+\mathcal{B}_{-1}\right\Vert _{\mathbf{Tr}}\leq\left\Vert \mathcal{B}_{+1}\right\Vert _{\mathbf{Tr}}+\left\Vert \mathcal{B}_{-1}\right\Vert _{\mathbf{Tr}},\]
thus\[
\left\Vert \mathcal{B}\right\Vert _{\mathbf{Tr}}\leq2\cdot\left\Vert \mathcal{B}_{\pm1}\right\Vert _{\mathbf{Tr}}<\infty.\]
The operator $\mathcal{B}$ is hence also of trace class. By the properties
of trace class operators \cite{Yafaev,TraceideaSimon,Kato} one can
define the convergent determinant $\det\left(\mathcal{A}\right)=\prod_{\kappa\in\mathbb{Z}}\lambda_{\kappa}\left(\mathcal{A}\right)$
of the operator $\mathbf{I}+\mathcal{B}=\mathcal{A}$ where $\lambda_{\kappa}\left(\mathcal{A}\right)$
are the eigenvalues of $\mathcal{A}$. The operator, by the Fredholm
alternatives, is then boundedly invertible if and only if $\det\left(\mathcal{A}\right)\neq0$
\cite{Yafaev}.

The recursion system of eq. (\ref{eq:recursionfourier}) will render
$\psi$ zero if two consecutive coefficients vanish. Seeking the non-trivial
solutions of eq. (\ref{eq:matrixequ}) is then equivalent to demanding
that \[
\det\left(\mathcal{A}\left(\nu;\, a,\, q\right)\right)=0.\]

\subsection{\label{sec:Whittaker-Hill-formula}Whittaker-Hill formula}

Whittaker was able to produce an astounding formula for the determinant
corresponding to the full Hill equation \cite{Whittaker}. We will
describe this result for the determinant $\Delta\left(a,\,\nu\right)=\det\left(\mathcal{A}\left(\nu;\, a,\, q\right)\right)$
for the Mathieu problem following the discourse of McLachlan \cite{McLachlan}.

Since $\kappa$ takes all values from $-\infty$ to $+\infty$, the
determinant $\Delta\left(a,\,\nu\right)$ will be left invariant under
the change $\nu\rightarrow-\nu$, hence $\Delta\left(a,\,\nu\right)$
is even in $\nu$, i.e., $\Delta\left(a,\,-\nu\right)=\Delta\left(a,\,\nu\right)$.
Furthermore, $\left(2\kappa-\left(\nu+2\right)\right)=\left(2\left(\kappa+1\right)-\nu\right)$,
giving $\Delta\left(a,\,\nu+1\right)=\Delta\left(a,\,\nu\right)$
again since $\kappa$ takes all values from $-\infty$ to $+\infty$,
i.e. the determinant is a periodic function in $\nu$ with period
$1$. We can hence restrict the study of the analytic behavior of
$\Delta\left(a,\,-\nu\right)$ to the strip $0\leq\nu\leq1$. $\Delta$
is a determinant and thus is a sum of products of $\xi_{i}\left(a,\,\nu\right)$.
Disregarding their poles, the functions $\xi_{i}\left(a,\,\nu\right)$ are
analytic functions hence so is $\Delta$ except at the poles of the functions
$\xi_{i}\left(a,\,\nu\right)$. We see by the structure of $\mathcal{A}$
that no term in $\xi_{2\kappa}$ will appear in higher powers than
$1$, i.e., all poles for the values of $a$ and $\nu$ satisfying eq.
(\ref{eq:constranu}) will be simple.

The function $\mathcal{D}\left(a,\,\nu\right)$,\[
\mathcal{D}\left(a,\,\nu\right)=\frac{1}{\cos\left(\pi\nu\right)-\cos\left(\pi\sqrt{a}\right)}\]
also has simple poles at those values of $a$ and $\nu$. If the function
$C\left(\nu\right)$ is appropriately chosen, the function\[
\Theta\left(a,\,\nu\right)=\Delta\left(a,\nu\right)-C\left(\nu\right)\cdot\mathcal{D}\left(a,\,\nu\right)\]
has no singularities. Note that $\Delta\left(a,\,\nu\right)$ has only one pole on the strip 
$0\leq\nu\leq1$. By the periodicity (in $\nu$) of $\Delta\left(a,\,\nu\right)$, finding the 
function $C\left(\nu\right)$ is reduced to calculating the constant $C=C\left(\nu\right)$ 
corresponding to the quotient between the residuals of the functions $\Delta\left(a,\,\nu\right)$ 
and $\mathcal{D}\left(a,\,\nu\right)$ at that pole. With the proper value of $C$, 
$\Theta\left(a,\,\nu\right)$ is analytic on the whole complex plane and has no poles. It must hence, 
by Liouville's theorem, be a constant.

In the limiting case $\nu\rightarrow+\imath\infty$, the matrix $\mathcal{A}\left(\nu;\, a,\, q\right)$
becomes diagonal dominant, i.e., only the diagonal elements (equal
to 1) remain in this limit since $\lim_{\nu\rightarrow\imath\infty}\xi_{2\kappa}=0,\,\forall\kappa\in\mathbb{Z}$.
Thus\[
\lim_{\nu\rightarrow+\imath\infty}\Delta\left(a,\,\nu\right)=1.\]
We see further that $\lim_{\nu\rightarrow\imath\infty}\mathcal{D}\left(a,\,\nu\right)=0$,
hence $\Theta\left(a,\,\nu\right)=1$. It follows that \begin{equation}
C=\frac{\Delta\left(a,\,\nu\right)-1}{\mathcal{D}\left(a,\,\nu\right)}.\label{eq:Whittaker1}\end{equation}
For the case $\nu=0$ we have $\mathcal{D}\left(a,\,0\right)=1/\left(1-\cos\left(\pi\sqrt{a}\right)\right)$.
Excluding the values of $a=4\kappa^{2}$ for which $\Delta\left(a,\,\nu=0\right)$
cannot be defined, we can write\[
C=\frac{\Delta\left(a,\,0\right)-1}{1-\cos\left(\pi\sqrt{a}\right)}.\]
By substitution into eq. (\ref{eq:Whittaker1}) one finds that \begin{equation}
\sin^{2}\left(\frac{\pi\nu}{2}\right)=\Delta\left(a,\,0\right)\sin^{2}\left(\frac{\pi\sqrt{a}}{2}\right),\, a\neq4\kappa^{2}.\label{WittakerHill}\end{equation}
The characteristic exponents resulting from the eigenvalue $a$ are
solutions of this equation. This formula was shown by Whittaker \cite{Whittaker}
for the full Hill problem. The full Hill problem of eq. (\ref{eq:Hill})
has an infinite number of Fourier coefficients. The matrix $\mathcal{A}$
will in this case be a full matrix, i.e., not merely a tridiagonal matrix
as in the Mathieu problem.

\section{\label{sec:Calculating}Calculating $\Delta\left(0\right)$}

\subsection{$\Delta\left(0\right)$ as the limiting case of a recursion}

By the Whittaker-Hill formula, the calculation of the characteristic
exponents of the eigenvalue $a$ is reduced to the calculation of
$\Delta\left(0\right)=\Delta\left(a,\,\nu=0\right)$. The determinant
of an infinite dimensional square matrix can be defined by recursion.
Defining the square tridiagonal matrices \[
\mathcal{A}_{i}^{0}=\left(\begin{array}{cccccccc}
1 & \xi_{2i}^{0} & 0\\
\xi_{2\left(i-1\right)}^{0} & 1 & \xi_{2\left(i-1\right)}^{0}\\
0 & \xi_{2\left(i-2\right)}^{0} & 1\\
 &  &  & \ddots\\
 &  &  &  & \ddots\\
\\ &  &  &  &  & 1 & \xi_{2\left(i-2\right)}^{0} & 0\\
 &  &  &  &  & \xi_{-2\left(i-1\right)}^{0} & 1 & \xi_{-2\left(i-1\right)}^{0}\\
 &  &  &  &  & 0 & \xi_{-2i}^{0} & 1\end{array}\right)\]
 with diagonal elements $1$. The off diagonal elements are collected
in the vectors $\left(\xi_{i-1}^{0},\,\xi_{i-2}^{0},\,\ldots,\,\xi_{0}^{0}\,,\ldots,\,\xi_{-i}^{0}\right)$
and $\left(\xi_{i}^{0},\,\ldots,\,\xi_{0}^{0}\,,\ldots,\,\xi_{-\left(i-2\right)}^{0},\,\xi_{-\left(i-1\right)}^{0}\right)$.
Here $\xi_{i}^{0}$ denotes $\xi_{i}\left(\nu=0\right)$ as previously
defined. The determinant $\Delta\left(0\right)$ is hence defined
as the limiting case of $\Delta_{i}^{0}=\det\left(\mathcal{A}_{i}^{0}\right)$,
i.e.,\[
\Delta\left(0\right)=\lim_{i\rightarrow\infty}\Delta_{i}^{0}.\]
Since for $\nu=0$, $\xi_{-i}^{0}=\xi_{i}^{0}$, we see that the matrices
$\mathcal{A}_{i}$ are symmetric with respect to the second diagonal.

We see that any matrix $\mathcal{A}_{i}^{0},\, i>2$, can be written
in terms of $\mathcal{A}_{i-1}^{0}$ through\[
\mathcal{A}_{i}^{0}=\left(\begin{array}{ccccc}
1 & \xi_{i}^{0}\\
\xi_{i-1}^{0} & \cdot & \cdot & \cdot\\
 & \cdot & \mathcal{A}_{i-1}^{0} & \cdot\\
 & \cdot & \cdot & \cdot & \xi_{i-1}^{0}\\
 &  &  & \xi_{i}^{0} & 1\end{array}\right),\, i>2,\]
where the dotted square represent the outer elements of $\mathcal{A}_{i-1}^{0}$.
Defining the truncation operators of a matrix $M$: $r\left(M\right)$,
$l\left(M\right)$, $u\left(M\right)$, $d\left(M\right)$, i.e.,
the matrices $M$ without left $\left(r\left(M\right)\right)$, respectively
right $\left(l\left(M\right)\right)$, columns and the upper $\left(d\left(M\right)\right)$,
respectively lower$\left(u\left(M\right)\right)$, rows, we can write
the Laplace decomposition of the determinant of the matrix $\mathcal{A}_{i}^{0}$.
Developing along the first row we get\begin{eqnarray*}
\det\left(\mathcal{A}_{i}^{0}\right) & = & \left|\begin{array}{cccc}
\cdot & \cdot & \cdot\\
\cdot & \mathcal{A}_{i-1}^{0} & \cdot\\
\cdot & \cdot & \cdot & \xi_{i-1}^{0}\\
 &  & \xi_{i}^{0} & 1\end{array}\right|-\\
 &  & \quad-\xi_{i}^{0}\left|\begin{array}{ccccc}
\xi_{i-1}^{0} & \cdot & \cdot & \cdot\\
 & \cdot & r\left(\mathcal{A}_{i-1}^{0}\right) & \cdot\\
 & \cdot & \cdot & \cdot & \xi_{i-1}^{0}\\
 &  &  & \xi_{i}^{0} & 1\end{array}\right|,\end{eqnarray*}
and then developing the first term on the last row and the second
term by the first column,\begin{eqnarray*}
\det\left(\mathcal{A}_{i}^{0}\right) & = & \det\left(\mathcal{A}_{i-1}^{0}\right)-\xi_{i}^{0}\left|\begin{array}{cccc}
\cdot & \cdot & \cdot\\
\cdot & l\left(\mathcal{A}_{i-1}^{0}\right) & \cdot\\
\cdot & \cdot & \cdot & \xi_{i-1}^{0}\end{array}\right|-\\
 &  & \quad-\xi_{i}^{0}\xi_{i-1}^{0}\left|\begin{array}{cccc}
\cdot & \cdot & \cdot\\
\cdot & dr\left(\mathcal{A}_{i-1}^{0}\right) & \cdot\\
\cdot & \cdot & \cdot & \xi_{i-1}^{0}\\
 &  & \xi_{i}^{0} & 1\end{array}\right|.\end{eqnarray*}
Proceeding in a similar fashion one can write\begin{eqnarray*}
\det\left(\mathcal{A}_{i}^{0}\right) & = & \Delta_{i-1}^{0}-\xi_{i}^{0}\xi_{i-1}^{0}\det\left(ul\left(\mathcal{A}_{i-1}^{0}\right)\right)-\xi_{i}^{0}\xi_{i-1}^{0}\det\left(rd\left(\mathcal{A}_{i-1}^{0}\right)\right)-\\
 &  & \quad\quad+\left(\xi_{i}^{0}\xi_{i-1}^{0}\right)^{2}\det\left(uldr\left(\mathcal{A}_{i-1}^{0}\right)\right).\end{eqnarray*}
One sees directly that $uldr\left(\mathcal{A}_{i-1}^{0}\right)=\mathcal{A}_{i-2}^{0}$.
Furthermore, by the symmetry properties around the second diagonal,
$\det\left(rd\left(\mathcal{A}_{i-1}^{0}\right)\right)=\det\left(ul\left(\mathcal{A}_{i-1}^{0}\right)\right)$.
Thus we conclude that\begin{equation}
\Delta_{i}^{0}=\Delta_{i-1}^{0}-2\cdot\xi_{i}^{0}\xi_{i-1}^{0}\det\left(rd\left(\mathcal{A}_{i-1}^{0}\right)\right)+\left(\xi_{i}^{0}\xi_{i-1}^{0}\right)^{2}\Delta_{i-2}^{0}.\label{eq:ietresc}\end{equation}
By simply rewriting eq. (\ref{eq:ietresc})\[
\Delta_{i}^{0}-\Delta_{i-1}^{0}=\left(\xi_{i}^{0}\xi_{i-1}^{0}\right)^{2}\Delta_{i-2}^{0}-2\cdot\xi_{i}^{0}\xi_{i-1}^{0}\det\left(rd\left(\mathcal{A}_{i-1}^{0}\right)\right),\]
we see that the series converges since $\lim_{i\rightarrow\infty}\xi_{i}=0$.

By further defining $\Omega_{i}=\det\left(ul\left(\mathcal{A}_{i}^{0}\right)\right)=\det\left(rd\left(\mathcal{A}_{i}^{0}\right)\right)$
and decomposing by minors, one can write\begin{equation}
\Omega_{i}=\det\left(\mathcal{A}_{i-1}^{0}\right)-\xi_{i}^{0}\xi_{i-1}^{0}\Omega_{i-1}.\label{eq:omegarec}\end{equation}
Substituting $\det\left(rd\left(\mathcal{A}_{i-1}^{0}\right)\right)$
in eq. (\ref{eq:ietresc}) one finds the recursion in $\Delta_{i}$,\begin{equation}
\Delta_{i}=\beta_{i}\Delta_{i-1}-\alpha_{i}\beta_{i}\Delta_{i-2}+\alpha_{i}\alpha_{i-1}^{2}\Delta_{i-3},\label{eq:iteration1}\end{equation}
where we have substituted $\xi_{i}^{0}\xi_{i-1}^{0}=\alpha_{i}$ and
$1-\alpha_{i}=\beta_{i}$.

\subsection{\label{recursol}A solution to linear third-order recursions without
constant terms}

We consider the recursion\begin{equation}
u_{i}=a_{i}u_{i-1}+b_{i}u_{i-2}+c_{i}u_{i-3}.\label{eq:rec1}\end{equation}
By simply rewriting it for $i-1$ and multiplicating it with $-c_{i}/b_{i-1}$,
assuming $b_{i}\neq0,\,\forall i$,\begin{equation}
-\frac{c_{i}}{b_{i-1}}u_{i-1}=-\frac{a_{i-1}c_{i}}{b_{i-1}}u_{i-2}-c_{i}u_{i-3}-\frac{c_{i}c_{i-1}}{b_{i-1}}u_{i-4}.\label{eq:rec2}\end{equation}
 Summing eqs. (\ref{eq:rec1}) and (\ref{eq:rec2}) one finds\begin{equation}
u_{i}=\left(a_{i}+\frac{c_{i}}{b_{i-1}}\right)u_{i-1}+\left(b_{i}-\frac{a_{i-1}c_{i}}{b_{i-1}}\right)u_{i-2}-\frac{c_{i}c_{i-1}}{b_{i-1}}u_{i-4}.\label{eq:firstrec}\end{equation}
One can hence, by this iteration procedure, construct a sum equal
to some $u_{i}$ over $n$ elements $u_{k}$, where the coefficient
for $u_{i-n+1}$ vanishes,\begin{equation}
u_{k}=\left[\sum_{j=1}^{n-2}\left\{ g_{n}^{k}\right\} _{k-j}u_{k-j}\right]+\left\{ \mathcal{G}_{n}^{k}\right\} u_{k-n}.\label{eq:recsum}\end{equation}
Here $\left\{ g_{n}^{k}\right\} _{k-j}$ are the coefficients corresponding
to the sum equal to $u_{k}$ with last term in the sum $u_{k-n}$.
Using this ansatz and eliminating $\left\{ \mathcal{G}_{n}^{k}\right\} u_{k-n}$
as previously one finds that\begin{eqnarray*}
\left\{ g_{n+1}^{k}\right\} _{k-j} & = & \left\{ \begin{array}{ll}
\lefteqn{\left\{ g_{n}^{k}\right\} _{k-j}} & ,\,1<i<n-3,\\
\\\left\{ g_{n}^{k}\right\} _{k-\left(n-2\right)}+\frac{\left\{ \mathcal{G}_{n}^{k}\right\} }{b_{k-\left(n-2\right)}} & ,\, j=n-2\\
\\-\left\{ \mathcal{G}_{n}^{k}\right\} \frac{a_{k-\left(n-2\right)}}{b_{k-\left(n-2\right)}}\left\{ g_{n}^{k}\right\} _{k-\left(n-1\right)} & ,\, j=n-1\end{array}\right.,\\
\left\{ \mathcal{G}_{n+1}^{k}\right\}  & = & -\left\{ \mathcal{G}_{n}^{k}\right\} \frac{c_{k-\left(n-2\right)}}{b_{k-\left(n-2\right)}}.\end{eqnarray*}
Since these recursions are of first order, one can deduce\begin{eqnarray*}
\left\{ g_{n+1}^{k}\right\} _{k-j} & = & \left\{ \begin{array}{ll}
\left\{ g_{n}^{k}\right\} _{k-j} & ,\,1<i<n-3\\
\\\left(-1\right)^{n-5}\cdot\left\{ g_{n}^{k}\right\} _{k-\left(n-2\right)}\times & ,\, j=n-2\\
\quad\times\frac{c_{k}c_{k-1}}{b_{k-\left(n-2\right)}\cdot b_{k-1}}\cdot\prod_{j=3}^{n-3}\frac{c_{k-j}}{b_{k-j}}\\
\\\left(-1\right)^{n-6}\cdot\frac{a_{k-\left(n-2\right)}\cdot c_{k}\cdot c_{k-1}}{b_{k-\left(n-2\right)}b_{k-1}}\times & ,\, j=n-1\\
\quad\times\prod_{j=3}^{n-3}\frac{c_{k-j}}{b_{k-j}}\\
\end{array}\right.,\\
\left\{ \mathcal{G}_{n+1}^{k}\right\}  & = & \left(-1\right)^{n-5}\cdot\frac{c_{k}c_{k-1}}{b_{k-1}}\cdot\prod_{j=3}^{n-3}\frac{c_{k-j}}{b_{k-j}},\end{eqnarray*}
from the initial values of the recursion given by eq. (\ref{eq:firstrec}).
We can hence write all $\left\{ g_{n}^{k}\right\} _{k-j}$ and $\left\{ \mathcal{G}_{n+1}^{k}\right\} $
explicitly. 

Being able to write such sums for any relevant $u_{k}$ and $n$,
one can eliminate the term of highest index on the r.h.s. of eq. (\ref{eq:recsum}).
If we consider

\begin{eqnarray*}
u_{k} & = & \left[\sum_{j=1}^{n-2}\left\{ g_{n}^{k}\right\} _{k-j}u_{k-j}\right]+\left\{ \mathcal{G}_{n}^{k}\right\} u_{k-n}\\
 & = & \left\{ g_{n}^{k}\right\} _{k-1}u_{k-1}+\left[\sum_{j=2}^{n-2}\left\{ g_{n}^{k}\right\} _{k-j}u_{k-j}\right]+\left\{ \mathcal{G}_{n}^{k}\right\} u_{k-n}\end{eqnarray*}
 and \begin{eqnarray*}
u_{k-1} & = & \left[\sum_{j=1}^{m-2}\left\{ g_{m}^{k-1}\right\} _{k-1-j}u_{k-j}\right]+\left\{ \mathcal{G}_{m}^{k-1}\right\} u_{k-m},\end{eqnarray*}
summing the two we find\begin{eqnarray*}
u_{k} & = & \left[\sum_{j=2}^{n-2}\left\{ g_{n}^{k}\right\} _{k-j}u_{k-j}\right]+\left\{ \mathcal{G}_{n}^{k}\right\} u_{k-n}+\left\{ g_{n}^{k}\right\} _{k-1}\left[\sum_{j=2}^{m-1}\left\{ g_{m}^{k-1}\right\} _{k-j}u_{k-j}\right]+\\
 &  & \quad+\left\{ g_{n}^{k}\right\} _{k-1}\left\{ \mathcal{G}_{m}^{k-1}\right\} u_{k-1-m}.\end{eqnarray*}
If we choose $m$ such that \[
m+1=n,\]
we find\begin{eqnarray}
u_{k} & = & \left(\left\{ g_{n}^{k}\right\} _{k-2}+\left\{ g_{n}^{k}\right\} _{k-1}\left\{ g_{n-1}^{k-1}\right\} _{k-2}\right)u_{k-2}+\nonumber \\
 &  & \quad+\left[\sum_{j=3}^{n-2}\left(\left\{ g_{n}^{k}\right\} _{k-j}+\left\{ g_{n}^{k}\right\} _{k-1}\left\{ g_{n-1}^{k-1}\right\} _{k-j}\right)u_{k-j}\right]+\nonumber \\
 &  & \quad+\left(\left\{ \mathcal{G}_{n}^{k}\right\} +\left\{ g_{n}^{k}\right\} _{k-1}\left\{ \mathcal{G}_{n-1}^{k-1}\right\} \right)u_{k-n}.\label{eq:firstel}\end{eqnarray}
Upon $\ell$ such eliminations, one can hence write $u_{k}$ as\begin{equation}
u_{k}=\sum_{j=\ell+1}^{n-2}\left\{ \gamma_{n}^{k}\right\} _{k-j}^{\ell}u_{k-j}+\left\{ \Gamma_{n}^{k}\right\} ^{\ell}u_{k-n}\label{eq:elansatz}\end{equation}
where $\left\{ \gamma_{n}^{k}\right\} _{k-j}^{\ell}$ and $\left\{ \Gamma_{n}^{k}\right\} ^{\ell}$
are the coefficients corresponding to $u_{k-j}$ and $u_{k-n}$ respectively,
after $\ell$ eliminations. If we proceed as previously and eliminate
$u_{k-\left(\ell+1\right)}$, we get\begin{eqnarray*}
u_{k} & = & \sum_{j=\ell+2}^{n-2}\left(\left\{ \gamma_{n}^{k}\right\} _{k-j}^{\ell}+\left\{ \gamma_{n}^{k}\right\} _{k-\ell-1}^{\ell}\left\{ g_{n-\ell}^{k-\ell}\right\} _{k-j}\right)u_{k-j}+\\
 &  & \quad+\left(\left\{ \gamma_{n}^{k}\right\} _{k-\ell-1}^{\ell}\left\{ \mathcal{G}_{n-\ell}^{k-\ell}\right\} +\left\{ \Gamma_{n}^{k}\right\} ^{\ell}\right)u_{k-n}.\end{eqnarray*}
Thus we can define recursions for $\left\{ \gamma_{n}^{k}\right\} _{k-j}^{\ell}$
and $\left\{ \Gamma_{n}^{k}\right\} ^{\ell}$ through\begin{eqnarray}
\left\{ \gamma_{n}^{k}\right\} _{k-j}^{\ell+1} & = & \left\{ \gamma_{n}^{k}\right\} _{k-j}^{\ell}+\left\{ \gamma_{n}^{k}\right\} _{k-\ell-1}^{\ell}\left\{ g_{n-\ell}^{k-\ell}\right\} _{k-j},\label{eq:recgamma}\\
\left\{ \Gamma_{n}^{k}\right\} ^{\ell+1} & = & \left\{ \gamma_{n}^{k}\right\} _{k-\ell-1}^{\ell}\left\{ \mathcal{G}_{n-\ell}^{k-\ell}\right\} +\left\{ \Gamma_{n}^{k}\right\} ^{\ell},\nonumber \end{eqnarray}
which are second, respectively first, order recurrences.

By calculating the coefficient of $u_{k-j}$ of highest index, i.e.,
$\left\{ \gamma_{n}^{k}\right\} _{k-\ell-1}^{\ell}$, one can recognize
a structure for these coefficients. $\left\{ \gamma_{n}^{k}\right\} _{k-\ell-1}^{\ell}$
can be written as a sum of products of $\left\{ g_{n-\mu}^{k-\mu}\right\} _{k-\nu}$
where the number of terms in the products range over all values from
$1$ to $\ell+1$. As an example, we consider such a sum for $\left\{ \gamma_{n}^{k}\right\} _{k-7}^{6}$
for terms with four $\left\{ g_{n-\mu}^{k-\mu}\right\} _{k-\nu}$
factors. These terms are written in the left column and the indices
$\left(\mu,\,\nu\right)$ corresponding to the terms $\left\{ g_{n-\mu}^{k-\mu}\right\} _{k-\nu}$
are shown in the right column,

\begin{eqnarray*}
\left\{ g_{n}^{k}\right\} _{k-1}\left\{ g_{n-1}^{k-1}\right\} _{k-2}\left\{ g_{n-2}^{k-2}\right\} _{k-3}\left\{ g_{n-3}^{k-3}\right\} _{k-6}+ & \quad\vdots\quad & \left(0,\,1\right)\left(1,\,2\right)\left(2,\,3\right)\left(3,\,6\right)\\
+\left\{ g_{n}^{k}\right\} _{k-1}\left\{ g_{n-1}^{k-1}\right\} _{k-2}\left\{ g_{n-2}^{k-2}\right\} _{k-4}\left\{ g_{n-4}^{k-4}\right\} _{k-6}+ & \quad\vdots\quad & \left(0,\,1\right)\left(1,\,2\right)\left(2,\,4\right)\left(4,\,6\right)\\
+\left\{ g_{n}^{k}\right\} _{k-1}\left\{ g_{n-1}^{k-1}\right\} _{k-2}\left\{ g_{n-2}^{k-2}\right\} _{k-5}\left\{ g_{n-5}^{k-5}\right\} _{k-6}+ & \quad\vdots\quad & \left(0,\,1\right)\left(1,\,2\right)\left(2,\,5\right)\left(5,\,6\right)\\
+\left\{ g_{n}^{k}\right\} _{k-1}\left\{ g_{n-1}^{k-1}\right\} _{k-3}\left\{ g_{n-3}^{k-3}\right\} _{k-4}\left\{ g_{n-4}^{k-4}\right\} _{k-6}+ & \quad\vdots\quad & \left(0,\,1\right)\left(1,\,3\right)\left(3,\,4\right)\left(4,\,6\right)\\
+\left\{ g_{n}^{k}\right\} _{k-1}\left\{ g_{n-1}^{k-1}\right\} _{k-3}\left\{ g_{n-3}^{k-3}\right\} _{k-5}\left\{ g_{n-5}^{k-5}\right\} _{k-6}+ & \quad\vdots\quad & \left(0,\,1\right)\left(1,\,3\right)\left(3,\,5\right)\left(5,\,6\right)\\
+\left\{ g_{n}^{k}\right\} _{k-1}\left\{ g_{n-1}^{k-1}\right\} _{k-4}\left\{ g_{n-4}^{k-4}\right\} _{k-5}\left\{ g_{n-5}^{k-5}\right\} _{k-6}+ & \quad\vdots\quad & \left(0,\,1\right)\left(1,\,4\right)\left(4,\,5\right)\left(5,\,6\right)\\
+\left\{ g_{n}^{k}\right\} _{k-2}\left\{ g_{n-2}^{k-2}\right\} _{k-3}\left\{ g_{n-3}^{k-3}\right\} _{k-4}\left\{ g_{n-4}^{k-4}\right\} _{k-6}+ & \quad\vdots\quad & \left(0,\,2\right)\left(2,\,3\right)\left(3,\,4\right)\left(4,\,6\right)\\
+\left\{ g_{n}^{k}\right\} _{k-2}\left\{ g_{n-2}^{k-2}\right\} _{k-3}\left\{ g_{n-3}^{k-3}\right\} _{k-5}\left\{ g_{n-5}^{k-5}\right\} _{k-6}+ & \quad\vdots\quad & \left(0,\,2\right)\left(2,\,3\right)\left(3,\,5\right)\left(5,\,6\right)\\
+\left\{ g_{n}^{k}\right\} _{k-2}\left\{ g_{n-2}^{k-2}\right\} _{k-4}\left\{ g_{n-4}^{k-4}\right\} _{k-5}\left\{ g_{n-5}^{k-5}\right\} _{k-6}+ & \quad\vdots\quad & \left(0,\,2\right)\left(2,\,4\right)\left(4,\,5\right)\left(5,\,6\right)\\
+\left\{ g_{n}^{k}\right\} _{k-3}\left\{ g_{n-3}^{k-3}\right\} _{k-4}\left\{ g_{n-4}^{k-4}\right\} _{k-5}\left\{ g_{n-5}^{k-5}\right\} _{k-6}\quad & \quad\vdots\quad & \left(0,\,3\right)\left(3,\,4\right)\left(4,\,5\right)\left(5,\,6\right).\end{eqnarray*}
One can construct the sequences of indices $\left(\mu,\,\nu\right)$
by considering a line with $\ell+1$ points labeled in ascending order
from $0$ to $\ell$. By further considering $p$ unidirectional non-stationary
jumps from the point labeled $0$ to the point labeled $\ell$, one
gets a sequence of initial and final points for the jumps e.g. $\left\{ \left(0,\,1\right)\left(1,\,2\right)\left(2,\,4\right)\left(4,\,6\right)\right\} $
for $\ell=6$ and $p=4$. We define the ensemble $\mathcal{S}_{p}^{\ell+1}$
of all such possible jumps, and by $\left\{ \mathcal{S}_{p}^{\ell+1}\right\} _{i}$
its $i$:th element. We can hence write the sum of all terms with
$p$ factors in $\left\{ g_{n-\mu}^{k-\mu}\right\} _{k-\nu}$ as\begin{equation}
\mathcal{P}_{p}^{\ell+1}=\sum_{i}\prod_{\left(\mu,\,\nu\right)\in\left\{ \mathcal{S}_{p}^{\ell+1}\right\} _{i}}\left\{ g_{n-\mu}^{k-\mu}\right\} _{k-\nu},\end{equation}
whereupon we find\begin{eqnarray}
\left\{ \gamma_{n}^{k}\right\} _{k-m-1}^{\ell} & = & \sum_{p=1}^{\ell+1}\mathcal{P}_{p}^{\ell+1}\\
 & = & \sum_{p=1}^{\ell+1}\sum_{i}\prod_{\left(\mu,\,\nu\right)\in\left\{ \mathcal{S}_{p}^{\ell+1}\right\} _{i}}\left\{ g_{n-\mu}^{k-\mu}\right\} _{k-\nu}.\nonumber \end{eqnarray}
With this, we can simply write the now first order recursions eqs.
(\ref{eq:recgamma}) as\begin{eqnarray}
\left\{ \gamma_{n}^{k}\right\} _{k-j}^{\ell+1} & = & \left\{ g_{n}^{k}\right\} _{k-j}+\sum_{\sigma=0}^{\ell}\left\{ \gamma_{n}^{k}\right\} _{k-m-1}^{\sigma}\cdot\left\{ g_{n-\sigma-1}^{k-\sigma-1}\right\} _{k-j},\\
\left\{ \Gamma_{n}^{k}\right\} ^{\ell+1} & = & \left\{ \mathcal{G}_{n}^{k}\right\} +\sum_{\sigma=0}^{\ell}\left\{ \gamma_{n}^{k}\right\} _{k-m-1}^{\sigma}\cdot\left\{ \mathcal{G}_{n-\sigma-1}^{k-\sigma-1}\right\} ,\nonumber \end{eqnarray}
or using the properties of $\left\{ \gamma_{n}^{k}\right\} _{k-m-1}^{\ell}$,\begin{eqnarray}
\left\{ \gamma_{n}^{k}\right\} _{k-j}^{\ell+1} & = & \left\{ g_{n}^{k}\right\} _{k-j}+\sum_{\sigma=0}^{\ell}\left[\sum_{p=1}^{\sigma+1}\mathcal{P}_{p}^{\ell+1}\right]\cdot\left\{ g_{n-\sigma-1}^{k-\sigma-1}\right\} _{k-j},\\
\left\{ \Gamma_{n}^{k}\right\} ^{\ell+1} & = & \left\{ \mathcal{G}_{n}^{k}\right\} +\sum_{\sigma=0}^{\ell}\left[\sum_{p=1}^{\sigma+1}\mathcal{P}_{p}^{\ell+1}\right]\cdot\left\{ \mathcal{G}_{n-\sigma-1}^{k-\sigma-1}\right\} .\nonumber \end{eqnarray}

The elimination process can be iterated until only three terms remain
in\begin{eqnarray}
u_{k} & = & \sum_{j=n-3}^{n-2}\left\{ \gamma_{n}^{k}\right\} _{k-j}^{n-4}u_{k-j}+\left\{ \Gamma_{n}^{k}\right\} ^{n-4}u_{k-n}\label{eq:solresur}\\
 & = & \left\{ \gamma_{n}^{k}\right\} _{k-n-3}^{n-4}u_{k-n-3}+\left\{ \gamma_{n}^{k}\right\} _{k-n-2}^{n-3}u_{k-n-2}+\left\{ \Gamma_{n}^{k}\right\} ^{n-4}u_{k-n}.\nonumber \end{eqnarray}
Here $\left\{ \gamma_{n}^{k}\right\} _{k-n-3}^{n-4}$, $\left\{ \gamma_{n}^{k}\right\} _{k-n-2}^{n-3}$
and $\left\{ \Gamma_{n}^{k}\right\} ^{n-4}$ can be calculated explicitly
with the help of $\left\{ g_{n-\mu}^{k-\mu}\right\} _{k-\nu}$. Since
$\left\{ g_{n-\mu}^{k-\mu}\right\} _{k-\nu}$ are also known explicitly,
eq. (\ref{eq:solresur}) is an explicit solution for $u_{k}$ knowing
$u_{k-n-3}$, $u_{k-n-2}$ and $u_{k-n}$ for some relevant $n$.

The only assumption we have made here is that $b_{i}\neq0,\,\forall i$.
Unfortunately eq. (\ref{eq:solresur}) is quite cumbersome. It is
hence difficult to make any general statement on the convergence of
eq. (\ref{eq:solresur}). By the simplicity of this scheme, we see
no apparent reason why it should not be possible to generalize to
any finite order recursion\begin{equation}
u_{i}=\sum_{j=1}^{k}a_{i-j}^{j}u_{i-j}.\end{equation}

\subsection{\label{sec:An-explicit-form}An explicit form for $\Delta\left(0\right)$}

Using the results of section \ref{recursol}, eq. (\ref{eq:iteration1})
can be written\begin{equation}
\Delta_{k}=\left\{ \gamma_{n}^{k}\right\} _{k-n-3}^{n-4}\Delta_{k-n-3}+\left\{ \gamma_{n}^{k}\right\} _{k-n-2}^{n-3}\Delta_{k-n-2}+\left\{ \Gamma_{n}^{k}\right\} ^{n-4}\Delta_{k-n}\label{eq:solution}\end{equation}
with \begin{eqnarray}
\left\{ \gamma_{n}^{k}\right\} _{k-j}^{\ell+1} & = & \left\{ g_{n}^{k}\right\} _{k-j}+\sum_{\sigma=0}^{\ell}\left[\sum_{p=1}^{\sigma+1}\mathcal{P}_{p}^{\ell+1}\right]\cdot\left\{ g_{n-\sigma-1}^{k-\sigma-1}\right\} _{k-j},\label{eq:gmnu1} \\
\nonumber\left\{ \Gamma_{n}^{k}\right\} ^{\ell+1} & = & \left\{ \mathcal{G}_{n}^{k}\right\} +\sum_{\sigma=0}^{\ell}\left[\sum_{p=1}^{\sigma+1}\mathcal{P}_{p}^{\ell+1}\right]\cdot\left\{ \mathcal{G}_{n-\sigma-1}^{k-\sigma-1}\right\} .\end{eqnarray}
Using the properties of the coefficients of eq. (\ref{eq:iteration1}),\begin{eqnarray}\label{eq:gmnu2}
\left\{ g_{n+1}^{k}\right\} _{k-j} & = & \left\{ \begin{array}{ll}
\left\{ g_{n}^{k}\right\} _{k-j} & ,\,1<j\leq n-3\\
\\\left(-1\right)^{n-5}\cdot\left\{ g_{n}^{k}\right\} _{k-\left(n-2\right)}\times & ,\, j=n-2\\
\quad\times\frac{\alpha_{k}\cdot\alpha_{k-1}^{3}\cdot\alpha_{k-2}^{2}}{\beta_{k-\left(n-2\right)}\cdot \beta_{k-1}}\cdot\prod_{j=3}^{n-3}\frac{\alpha_{k-j-1}^{2}}{\beta_{k-j}}\\
\\\left(-1\right)^{n-6}\cdot\frac{\alpha_{k}\cdot\alpha_{k-1}^{2}\cdot\alpha_{k-2}^{2}}{\alpha_{k-\left(n-2\right)}\beta_{k-1}}\times & ,\, j=n-1\\
\quad\times\prod_{j=3}^{n-3}\frac{\alpha_{k-j-1}^{2}}{\beta_{k-j}}\end{array}\right.,\\
\left\{ \mathcal{G}_{n+1}^{k}\right\}  & = & \left(-1\right)^{n-5}\cdot\frac{\alpha_{k}\cdot\alpha_{k-1}^{2}\cdot\alpha_{k-2}^{2}}{\beta_{k-1}}\cdot\prod_{j=3}^{n-3}\frac{\alpha_{k-j-1}^{2}}{\beta_{k-j}}\label{eq:gmnu3}.\end{eqnarray}
We have thus found an explicit expression for $\Delta\left(0\right)$.
Note that the denominators diverge for those values of $a$ for which
$\beta_{i}=1-\alpha_{i}=0,\,\forall i\in\mathbb{Z}$.

Unfortunately the form of eq. (\ref{eq:solution}) does not, \`{a}
priori, shed light on specific properties of the solutions of
the Whittaker-Hill formula eq. (\ref{WittakerHill}).

\subsection{Numerical considerations}

Considering the minors of determinants equivalent to $\Delta_{n}$, R. Sips was able to 
produce a method for calculating the characteristic exponent \cite{Sips}. His method effectively 
produces first order linear recurrences for $n$ minors $D_{i}$ with coefficients $\alpha_{i}$, 
whereupon the determinant $\Delta_{n}$ can be written 
\begin{equation} \nonumber \Delta_{n}=D_{1}^{2}-\alpha_{1}^{2}D_{2}^{2}.\end{equation} 
For small values of $q$, his method converges rapidly since $\alpha_{i}=\xi_{i}\xi_{i-1}$ 
decreases rapidly with growing index $i$. Having calculated a $\Delta_{n}$ one has, however, 
to calculate $\Delta_{n'}$ of higher order, to repeat the whole procedure.

We see that the convergence of our recurrence eq. (\ref{eq:iteration1}) 
\begin{equation}\nonumber
\Delta_{i}=\beta_{i}\Delta_{i-1}-\alpha_{i}\beta_{i}\Delta_{i-2}+\alpha_{i}\alpha_{i-1}^{2}\Delta_{i-3},\end{equation}
is also directly linked to the behavior of $\alpha_{i}$ $\left(\beta_{i}=1-\alpha_{i}\right)$. The cost of further calculating determinants of higher orders will, however, be considerably less than for the method of R. Sips, since one only has to successively iterate eq. (\ref{eq:iteration1}).

\section{Conclusion}

With the Whittaker-Hill formula we have derived a third order determinantal
recursion. By the decomposition of the third order recursion to first
order recursions we can write any relevant term of the recursion explicitly.
We believe that this scheme can be generalized to any linear recursion
of order greater than one. The explicit formula can be used to write
the sought after determinant explicitly. We expect applications to the 
numerical treatment of the relation between eigenvalues and characteristic 
exponents.
\begin{acknowledgement*}
I would like to thank Jens Bolte for valuable comments and discussions.
\end{acknowledgement*}

\end{document}